\documentclass{jnmp}

\usepackage{amsmath}

\JNMPnumberwithin{equation}{section}

\begin{document}

\renewcommand{\evenhead}{M~A~Jafarizadeh and S~Behnia}
\renewcommand{\oddhead}{Hierarchy of Chaotic Maps with an Invariant Measure}

\thispagestyle{empty}

\FirstPageHead{9}{1}{2002}{\pageref{Jafarizadeh-firstpage}--\pageref{Jafarizadeh-lastpage}}{Article}

\copyrightnote{2002}{M~A~Jafarizadeh and S~Behnia}

\Name{Hierarchy of Chaotic Maps with an Invariant Measure and their
Compositions}

\label{Jafarizadeh-firstpage}

\Author{M~A~JAFARIZADEH~${}^{\dag^{1}}{}^{\dag^{2}}{}^{\dag^{3}}$
and S~BEHNIA~${}^{\dag^{2}}{}^{\dag^{4}}$}

\Address{${}^{\dag^{1}}$~Department of Theoretical
Physics and Astrophysics, Tabriz University,\\
$\phantom{{}^{\dag^{1}}}$~Tabriz 51664, Iran\\
$\phantom{{}^{\dag^{1}}}$~E-mail: jafarzadeh@ark.tabrizu.ac.ir\\[10pt]
${}^{\dag^{2}}$~Institute for Studies in Theoretical Physics and
Mathematics,\\
$\phantom{{}^{\dag^{1}}}$~Teheran 19395-1795, Iran\\[10pt]
${}^{\dag^{3}}$~Pure and Applied Science Research Center, Tabriz 51664, Iran\\[10pt]
${}^{\dag^{4}}$~Department of Physics, IAU, Urmia, Iran}

\Date{Received February 12, 2001; Revised (1) August 22, 2001;
Revised (2) September 29, 2001;
Accepted October 01, 2001}

\begin{abstract}
\noindent We give a hierarchy of many-parameter families of maps
of the interval $[0,1]$ with an invariant measure and using the
measure, we calculate Kolmogorov--Sinai entropy of these maps
analytically. In contrary to the usual one-dimensional maps these
maps do not possess period doubling or period-$n$-tupling cascade
bifurcation to chaos, but they have single fixed point attractor
at certain region of parameters space, where they bifurcate
directly to chaos without having period-$n$-tupling scenario
exactly at certain values of the parameters.
%\\\\ {\bf Keywords:Chaos, Invariant measure, Kolmogorov-Sinai entropy, chaotic dynamical
%systems }.\\
%{\bf PACs numbers:05.45.Ra, 05.45.Jn, 05.45.Tp }
\end{abstract}

\section{Introduction}
In the past twenty years dynamical systems, particularly one
dimensional iterative maps have attracted much attention and have
become an important area of research activity. One of the
landmarks in it was introduction of the concept of
Sinai--Ruelle--Bowen (SRB) measure or natural invariant measure
\cite{sinai, jakob}. This is, roughly speaking, a measure that is
supported on an attractor and also describes statistics of long
time behavior of the orbits for almost every initial condition in
 corresponding basin of attractor. This measure can be obtained
by computing the fixed points of the so called
Frobenius--Perron~(FP) operator which can be viewed as a
differential-integral operator.The exact determination of
invariant measure of dynamical systems is rather a nontrivial task
taking into account that the invariant measure of few dynamical
systems such as one-parameter family of one-dimensional piecewise
linear maps~\cite{tasaki,tasaki1, Umeno} including Baker and tent
maps or unimodal maps such as logistic map for certain values of
its parameter can be derived analytically. In most  cases only
numerical algorithms, as an example Ulam's method
\cite{froy,froy1,blank}, are used for computation of fixed points
of FP-operator.

Here in this paper we give a new hierarchy of many-parameter
families of maps of the interval $[0,1]$ with an invariant
measure. These maps are defined as ratios of polynomials where we
have derived analytically their invariant measure for arbitrary
values of the parameters. Using this measure, we have calculated
analytically Kolomogorov--Sinai~(KS) entropy of these maps. It is
shown that they possess very peculiar property, that is, contrary
to the usual maps, these do not possess period doubling or
period-$n$-tupling cascade bifurcation to chaos, but instead they
have single fixed point attractor at certain region of parameters
space where they bifurcate directly to chaos without having
period-$n$-tupling scenario.

 The paper is organized as
follows: In Section~2 we introduce the hierarchy of
many-parameter families of chaotic maps. In Section~3 we show
that the invariant measure is actually the eigenstate of the
FP-operator corresponding to largest eigenvalue~$1$. Then in
section IV using this measure we calculate KS-entropy of these
maps for an arbitrary values of the parameters. Paper ends with a
brief conclusion.

\section{Hierarchy of chaotic maps with an invariant measure\\
 and their compositions}
 Let us first consider the one-parameter families of
chaotic maps of the interval $[0,1]$ defined as the ratios of
polynomials of degree $N$:
\begin{gather}
\Phi_{N}(x,\alpha)=\frac{\alpha^2\left(1+(-1)^N{
}_2F_1\left(-N,N,\frac{1}{2},x\right)\right)}
{(\alpha^2+1)+(\alpha^2-1)(-1)^N{ }_2F_1\left(-N,N,\frac{1}{2},x\right)}\nonumber\\
\phantom{\Phi_{N}(x,\alpha)} {}=\frac{\alpha^2(T_N(\sqrt{x}))^{2}}{1+(\alpha^2-1)(T_N(\sqrt{x})^{2})},
  \end{gather}
where $N$ is an integer greater than $1$. Also
\[
_2F_1\left(-N,N,\frac{1}{2},x\right)=(-1)^{N}\cos{\left(2N\arccos\sqrt{x}\right)}=(-1)^{N}T_{2N}(\sqrt{x})
\]
 are hypergeometric polynomials of degree $N$ and $T_{N}(x)$ are
Chebyshev polynomials of type I~\cite{wang} respectively.
Obviously these maps unit interval $[0,1]$ into itself.
 $ \Phi_{N}(x,\alpha)$ is $(N-1)$-nodal map, that is it has
 $(N-1)$ critical points in unit interval $[0,1]$, since its
 derivative is proportional to derivative of hypergeometric
 polynomial $_2F_1\left(-N,N,\frac{1}{2},x\right)$ which is itself a hypergeometric
 polynomial of degree $(N-1)$, hence it has
 $(N-1)$ real roots in unit interval $[0,1]$. Defining Shwarzian
 derivative \cite{dev} ${S\Phi_N(x,\alpha)}$ as:
\[
S\left(\Phi_N(x,\alpha)\right)=\frac{\Phi_{N}^{\prime\prime\prime}
(x,\alpha)}{\Phi_{N}^{\prime}(x,\alpha)}-\frac{3}{2}\left(\frac{\Phi_{N}^{\prime\prime}(x,\alpha)}{\Phi_{N}^{\prime}(x,\alpha)}\right)^2=\left(\frac{\Phi_{N}^{\prime\prime}(x,\alpha)}{\Phi_{N}^{\prime}(x,\alpha)}\right)^{\prime}-\frac{1}{2}
\left(\frac{\Phi_{N}^{\prime\prime}(x,\alpha)}{\Phi_{N}^{\prime}(x,\alpha)}\right)^2,
\]
 with a prime denoting differentiation with respect to variable $x$, one can show that
 (see Appendix~A):
\[
S\left(\Phi_{N}(x,\alpha)\right)=S\left(_2F_1\left(-N,N,\frac{1}{2},x\right)\right)\leq0.
\]
 Therefore, the maps $\Phi_{N}^{\alpha}(x)$ have at most $N+1$
attracting periodic orbits~\cite{dev}. Using the above hierarchy
of family of one-parameter chaotic maps, we can generate new
hierarchy of families of many-parameter chaotic maps with an
invariant measure simply from the composition of these maps. Hence
considering the functions $\Phi_{N_{k}}(x,\alpha_k)$,
$k=1,2,\ldots,n$ we denote their composition by:
$\Phi_{N_{1},N_{2},\ldots,N_{n}}^{\alpha_1,\alpha_2,\ldots,\alpha_n}(x)$
which can be written in terms of  them in the following form:
\begin{gather}
\nonumber\Phi_{N_1,N_2,\ldots,N_n}^{\alpha_1,\alpha_2,\ldots,\alpha_n}(x)=
\overbrace{\left(\Phi_{N_1}\circ\Phi_{N_2}\circ\ldots\circ\Phi_{N_n}(x)\right)}^n
\\
\phantom{\Phi_{N_1,N_2,\ldots,N_n}^{\alpha_1,\alpha_2,\ldots,\alpha_n}(x)}
=\Phi_{N_1}\left(\Phi_{N_2}(\ldots(\Phi_{N_n}(x,\alpha_n),\alpha_{(n-1)})\ldots,\alpha_2),
\alpha_1\right).
\end{gather}
Since these maps consist of composition of $(N_{k}-1)$-nodals
$(k=1,2,\ldots,n)$ maps with negative Shwarzian derivative, they
are $(N_1N_2 \cdots N_{n}-1)$-nodals maps and their Shwarzian
derivative is negative, too \cite{dev}. Therefore these maps  have
at most $ N_1N_2 \cdots N_n+1$ attracting periodic
orbits~\cite{dev}. As it is shown below in this section, these
maps have only a single period one stable fixed points. Denoting
m-composition of these functions by~$ \Phi^{(m)} $, it is
straightforward to show that the derivative of $ \Phi^{(m)} $ at
its possible $ m\times n $ periodic points of an $m$-cycle can be
defined as: $x_{\mu,k+1}=\Phi_{N_k}(x_{\mu,k},\alpha_{k})$,
$x_{1,\mu+1} =\Phi_{N_n}(x_{n,\mu},\alpha_N)$, and
$x_{1,1}=\Phi_{N_n}(x_{m,n},\alpha_n)$, $\mu=1,2\ldots,m$,
$k=1,2,\ldots,n $  is
\begin{equation}
\left|\frac{d}{dx}\Phi^{(m)}\right|
=\prod_{\mu=1}^{m}\prod_{k=1}^{n}
\left|\frac{N_k}{\alpha_k}\left(\alpha_k^{2}+\left(1-\alpha_k^{2}\right)x_{\mu,k}\right)\right|,
\end{equation}
since for $x_{\mu,k}\in [0,1]$ we have:
\[
\min\left(\alpha_k^{2}+\left(1-\alpha_k^{2}x_{\mu,k}\right)\right)=\min\left(1,\alpha_k^{2}\right),
\]
therefore:
\[
\min\left|\frac{d}{dx}\Phi^{(m)}\right|=
\prod_{k=1}^{n}\left(\frac{N_k}{\alpha_k}\min\left(1,\alpha_k^{2}\right)\right)^{m}.
\]
Hence, the above expression is definitely  greater than $1$ for $
\prod\limits_{k=1}^{n}\frac{1}{N_k}<
\prod\limits_{k=1}^{n}\alpha_k <\prod\limits_{k=1}^{n} N_k $, that
is, these maps do not have any kind of $m$-cycle or periodic
orbits in the region of the parameters space defined by
$\prod\limits_{k=1}^{n} \frac{1}{N_k}<\prod\limits_{k=1}^{n}
\alpha_k <\prod\limits_{k=1}^{n} N_k $, actually they are chaotic
in this region of the  parameters space. From (2.3) it follows
that $\left|\frac{d}{dx}\Phi^{(m)}\right|$ at $ m \times n $
periodic points of $m$-cycle belonging to interval $[0,1]$, vary
between $\prod\limits_{k=1}^{n}{\left(N_k\alpha_k\right)}^{m}$ and
$\prod\limits_{k=1}^{n}{\left(\frac{N_k}{\alpha_k}\right)}^{m}$
for
$\prod\limits_{k=1}^{n}\alpha_k<\prod\limits_{k=1}^{n}\frac{1}{N_k}$
and between
$\prod\limits_{k=1}^{n}\left(\frac{N_k}{\alpha_k}\right)^{m}$ and
$\prod\limits_{k=1}^{n}{\left(N_k\alpha_k\right)}^{m}$ for
$\prod\limits_{k=1}^{n}\alpha_k>\prod\limits_{k=1}^{n}N_k$,
respectively.

Definitely from the definition of these maps, we see that $x=1$
and $x=0$ (in special case of odd integer values of
$N-1,N_2,\ldots,N_n$ ) belong to one of $m$-cycles.

For $\prod\limits_{k=1}^{n}\alpha_k<\prod_{k=1}^{n}\frac{1}{N_k}$
$\left(\prod\limits_{k=1}^{n}\alpha_k>\prod\limits_{k=1}^{n}N_k\right)$,
the formula (2.3) implies that for those cases in which $x=1$
$(x=0)$ belongs to one of m-cycles, we have
$\left|\frac{d}{dx}\Phi^{(m)}\right|<1$, hence the curve of
$\Phi^{(m)}$ starts at $x=1$ $(x=0)$ beneath the bisector and then
crosses it at the previous (next) periodic point with slope
greater than one, since the formula (2.3) implies that the slope
of fixed points increases with the decreasing (increasing) of
$|x_{\mu,k}|$, therefore at all periodic points of n-cycles except
for $x=1$ $(x=0)$ the slope is greater than one that is they are
unstable, this is possible only if $x=1$ $(x=0)$ is the only
period one fixed point of these maps.

 Hence, all $m$-cycles except
for possible period one fixed points $x=1$ and $x=0$ are unstable.

 Actually, the fixed point $ x=0 $ is the stable fixed point of
these maps in the regions of the parameters spaces defined by
$\alpha_k>0$, $k=1,2,\ldots,n$ and
$\prod\limits_{k=1}^{n}\alpha_k<\prod\limits_{k=1}^{n}\frac{1}{N_k}$
only for odd integer values of $N_1,N_2,\ldots,N_n$, however, if
one of the integers $N_k$, $k=1,2,\ldots,n$ happens to be even,
then the $x=0$ will not  be a stable fixed point anymore. But the
fixed point $ x=1 $ is stable fixed point of these maps in the
regions of the parameters spaces defined by
$\prod\limits_{k=1}^{n}\alpha_k>\prod\limits_{k=1}^{n}N_k$ and
$\alpha_k<\infty$, $k=1,2,\ldots,n$ for all integer values of
$N_1,N_2,\ldots,N_n$.

 As an example we give below some of these
maps:
 \begin{gather}
\Phi_{2}^{\alpha}(x)=\frac{\alpha^{2}(2x-1)^{2}}{4x(1-x)+\alpha^{2}(2x-1)^{2}},\\
\Phi_{3}^{\alpha}(x)=\frac{\alpha^{2}x(4x-3)^{2}}{\alpha^{2}x(4x-3)^{2}+(1-x)(4x-1)^{2}},\\
\Phi_{4}^{\alpha}(x)=\frac{\alpha^{2}\left(1-8x(1-x)\right)^{2}}
{\alpha^2(1-8x(1-x))^{2}+16x(1-x)(1-2x)^{2}},\\
\Phi_{5}^{\alpha}(x)=\frac{\alpha^{2}x\left(16x^{2}-20x+5\right
)^{2}}{\alpha^{2}x(16x^{2}-20x+5)^{2}+(1-x)(16x^{2}-(2x-1))},\\
 \Phi_{2,2}^{\alpha_{1},\alpha_{2}}(x) ={\frac {{\alpha_{{1}}}^{2}\left
(4\,x\left (x-1\right )+\left (2\,x-1 \right
)^{2}{\alpha_{{2}}}^{2}\right )^{2}}{{\alpha_{{1}}}^{2}\left
(4\,x\left (x-1\right )+\left (2\,x-1\right
)^{2}{\alpha_{{2}}}^{2} \right
)^{2}-16x\alpha_{2}^{2}(2x-1)^{2}(x-1)}},
\\
\Phi_{2,3}^{\alpha_{1},\alpha_{2}}(x) =
{{\alpha_{1}}^{2}\left(\left(x-1\right)\left(4\,x-1\right)^{2}
+x\left(4\,x-3\right)^{2}{\alpha_{2}}^{2}\right)^{2}}\nonumber\\
\phantom{\Phi_{2,3}^{\alpha_{1},\alpha_{2}}(x) =}{}\times \Big(
{\alpha_{1}}^{2}\left(\left(x-1\right)\left(4\,x-1\right)^{2}+x\left(4\,x-3\right)^{2}
{\alpha_{{2}}}^{2}\right)^{2}\nonumber\\
\phantom{\Phi_{2,3}^{\alpha_{1},\alpha_{2}}(x) =}{}
-4x\alpha_{2}^{2}(x-1)(4x-1)^{2}(4x-3)^{2}\Big)^{-1},
\\
\Phi_{{3,2}}^{\alpha_{1},\alpha_{2}}(x)={{\alpha_{2}}^{2}\left(
(x-1 )\left (4\,x-1\right )^{2}+x \left (4\,x-3\right
)^{2}{\alpha_{1}}^{2}\right )^{2}}\nonumber\\
\phantom{\Phi_{{3,2}}^{\alpha_{1},\alpha_{2}}(x)=}{}\times\Big({\alpha_{{2}
}}^{2}\left (\left (x-1\right )\left (4\,x-1\right )^{2}+x\left
(4\,x -3\right )^{2}{\alpha_{{1}}}^{2}\right )^{2}\nonumber\\
\phantom{\Phi_{{3,2}}^{\alpha_{1},\alpha_{2}}(x)=}{}
+4x\alpha_{1}^{2}(x-1)(4x-1)^{2}(4x-3)^{2}\Big)^{-1},
\end{gather}
\begin{gather}
 \Phi_{{3,3}}^{\alpha_{1},\alpha_{2}}(x)={{\alpha_{1}}^{2}
 {\alpha_{2}}^{2}x\left (4\,x-3\right )^{2}\left
(3 \,\left (x-1\right )\left (4\,x-1\right )^{2}+x\left
(4\,x-3\right )^{2}{\alpha_{2}}^{2}\right )}\nonumber\\
\phantom{\Phi_{{3,3}}^{\alpha_{1},\alpha_{2}}(x)=}{}\times\Big(-\left
(x-1\right )^{3}\left (4\,x-1 \right )^{6}+3\,x\left
(3\,{\alpha_{1}}^{2}-2\right )\left (4\,x-3 \right
)^{2}\nonumber\\
\phantom{\Phi_{{3,3}}^{\alpha_{1},\alpha_{2}}(x)=}{}\times \left
(x-1\right )^{2}\left (4\,x-1\right
)^{4}{\alpha_{2}}^{2}+h\Big)^{-1},
\end{gather}
 where
\[
h=3x^{2}\alpha_{2}^{4}\left(-3+2\alpha^{2}_{1}\right)(x-1)(4x-1)^{2}(4x-3)^{4}
+\alpha_{1}^{2}\alpha_{2}^{6}x^{3}(4x-3)^{6}.
\]

  Below we also introduce their
conjugate or isomorphic maps which will be very useful in
derivation of their invariant measure and calculation of their
KS-entropy in the next section. Conjugacy means that the
invertible map $ h(x)=\frac{1-x}{x} $ maps $I=[0,1]$ into $
[0,\infty) $ and transforms  maps ${\Phi}_{N_k}(x,\alpha_k)$ into
$\tilde{\Phi}_{N_k}(x,\alpha_k)$ defined as:
\[
\tilde{\Phi}_{N_k}(x,\alpha_k)=h\circ\Phi_{N_k}(x,\alpha_k)\circ
h^{(-1)}=\frac{1}{\alpha_k^{2}}\tan^{2}\left(N_k\arctan\sqrt{x}\right),
\]
Hence, this transforms the maps
$\Phi_{N_{1},N_{2},\ldots,N_{n}}^{\alpha_1,\alpha_2,\ldots,\alpha_n}(x)$
into
$\tilde{\Phi}_{N_{1},N_{2},\ldots,N_{n}}^{\alpha_1,\alpha_2,\ldots,\alpha_n}(x)$
defined as:
\begin{gather}\hspace*{-27pt}
\tilde{\Phi}_{N_1,N_2,\ldots,N_n}^{\alpha_1,\alpha_2,\ldots,\alpha_n}(x)\nonumber\\
\hspace*{-27pt}=\frac{1}{\alpha_1^2}\tan^{2}\left(N_1\arctan\sqrt\circ
\frac{1}{\alpha_2^2}\tan^{2}\left(N_2\arctan\sqrt\circ\, \cdots\,
\circ\ \frac{1}{\alpha_n^2}\tan^{2}\left(N_n\arctan\sqrt
x\right)\right)\right)\nonumber\\
\hspace*{-27pt}=\frac{1}{\alpha_1^2}\tan^{2}\!\left(N_1\arctan\sqrt{
\frac{1}{\alpha_2^2}\tan^{2}\left(N_2\arctan\sqrt{\cdots
\frac{1}{\alpha_n^2}\tan^{2}\left(N_n\arctan\sqrt{x}\right)}\cdots\!\right)}\,\right).
\end{gather}

\section{Invariant measure}

 Dynamical systems, even apparently simple
dynamical systems such as maps of an interval, can display a rich
variety of different asymptotic behaviors. On measure theoretical
level these types of behavior are described by SRB \cite{sinai,
Dorf2} or invariant measure describing statistically stationary
states of the system. The probability measure $\mu$  on $[0,1]$ is
called an SRB or invariant measure of the maps
$\Phi_{N_{1},N_{2},\ldots,N_{n}}^{\alpha_1,\alpha_2,\ldots,\alpha_n}(x)$
 given
in $(2.2)$, if it is
$\Phi_{N_{1},N_{2},\ldots,N_{n}}^{\alpha_1,\alpha_2,\ldots,\alpha_n}(x)$-invariant
and absolutely continuous with respect to Lebesgue measure. For
deterministic system such as these composed maps, the
$\Phi_{N_{1},N_{2},\ldots,N_{n}}^{\alpha_1,\alpha_2,\ldots,\alpha_n}(x)$-invariance
 means that its invariant measure $\mu_{\Phi_{N_{1},N_{2},\ldots,N_{n}}^
 {\alpha_1,\alpha_2,\ldots,\alpha_n}}(x)$ fulfills the following
formal FP-integral equation:
\[
\mu_{\Phi_{N_{1},N_{2},\ldots,N_{n}}^{\alpha_1,\alpha_2,\ldots,\alpha_n}}(y)
=\int_{0}^{1}\delta\left(y-\Phi_{N_{1},N_{2},\ldots,N_{n}}^{\alpha_1,\alpha_2,\ldots,\alpha_n}(x)
\right)\mu_{\Phi_{N_{1},N_{2},\ldots,N_{n}}^{\alpha_1,\alpha_2,\ldots,\alpha_n}}(x)dx.
\]
This is equivalent to:
\begin{equation}
\mu_{\Phi_{N_{1},N_{2},\ldots,N_{n}}^{\alpha_1,\alpha_2,\ldots,\alpha_n}}(y)=\sum_{x\in\
\Phi_{N_{1},N_{2},\ldots,N_{n}}^{\alpha_1,\alpha_2,\ldots,\alpha_n}(y)
}\mu_{\Phi_{N_{1},N_{2},\ldots,N_{n}}^{\alpha_1,\alpha_2,\ldots,\alpha_n}}(x)\frac{dx}{dy},
\end{equation}
defining the action of standard FP-operator for the map $\Phi(x)$
over a function as:
\begin{equation}
P_{\Phi_{N_{1},N_{2},\ldots,
N_{n}}^{\alpha_{1},\alpha_{2},\ldots,\alpha_{n}} }f(y)=\sum_{x\in
\Phi_{N_{1},N_{2},\ldots,N_{n}}^{\alpha_{1},\alpha_{2},\ldots,\alpha_{n}}
(y)}f(x)\frac{dx}{dy}.
\end{equation}
We see that, the invariant measure
$\mu_{\Phi_{N_{1},N_{2},\ldots,N_{n}}^{\alpha_1,\alpha_2,\ldots,\alpha_n}}(x)$
is the eigenstate of the FP-operator
$P_{\Phi_{N_{1},N_{2},\ldots,N_{n}}^{\alpha_1,\alpha_2,\ldots,\alpha_n}}$
corresponding to the largest eigenvalue~1.

 As we will prove below,
the measure
$\mu_{\Phi_{N_{1},N_{2},\ldots,N_{n}}^{\alpha_1,\alpha_2,\ldots,\alpha_n}
}(x,\beta)$ defined as:
\begin{equation}
\frac{1}{\pi}\frac{\sqrt{\beta}}{\sqrt{x(1-x)}(\beta+(1-\beta)x)},
\end{equation}
is the invariant measure of the maps
$\Phi_{N_{1},N_{2},\ldots,N_{n}}^{\alpha_1,\alpha_2,\ldots,\alpha_n}(x)$
 provided that the parameter $\beta$ is positive and fulfills the following relation:
\begin{gather}
\prod _{k=1}^{n}\alpha_k
\times\frac{A_{N_{n}}\left(\frac{1}{\beta}\right)}{B_{N_{n}}\left(\frac{1}{\beta}\right)}\times
\frac{A_{N_{n-1}}\Big(\mbox{\raisebox{1.5mm}{$\frac{1}{\eta_{N_{n}}^{\alpha_{n}}
\left(\frac{1}{\beta}\right)}$}}\Big)}
{B_{N_{n-1}}\Big(\mbox{\raisebox{1.5mm}{$\frac{1}{\eta_{N_{n}}^{\alpha_{n}}
\left(\frac{1}{\beta}\right)}$}}\Big)} \nonumber\\ \qquad {}\times
\frac{A_{N_{n-2}}\Big(\mbox{\raisebox{1.5mm}{$\frac{1}{\eta_{N_{n-1},N_{n}}^
{\alpha_{n-1},\alpha_{n}}\left(\frac{1}{\beta}\right)}$}}\Big)}
{B_{N_{n-2}}\Big(\mbox{\raisebox{1.5mm}{$\frac{1}{\eta_{N_{n-1},N_{n}}^
{\alpha_{n-1},\alpha_{n}}\left(\frac{1}{\beta}\right)}$}}\Big)}
\times\cdot\times
\frac{A_{N_{1}}\Big(\mbox{\raisebox{1.5mm}{$\frac{1}{\eta_{N_{2},N_{3},\ldots,N_{n}}^
{\alpha_{2},\alpha_{3},\ldots,a_{n}}\left(\frac{1}{\beta}\right)}$}}\Big)}
{B_{N_{1}}\Big(\mbox{\raisebox{1.5mm}{$\frac{1}{\eta_{N_{2},N_{3},\ldots,N_{n}}^
{\alpha_{2},\alpha_{3},\ldots,\alpha_{n}}\left(\frac{1}{\beta}\right)}$}}\Big)}
=1,
\end{gather}
where the polynomials $ A_{N_{k}}(x)$ and $ B_{N_{k}}(x)$
$(k=1,2,\ldots,n)$ are defined as:
 \begin{gather}
 A_{N_{k}}(x) =\sum_{l=0}^{\left[ \frac{N_k}{2}\right]}C_{2l}^{N_k}x^{l},
\\
 B_{N_{k}}(x) =\sum_{l=0}^{\left[ \frac{N_k-1}{2}\right]}C_{2l+1}^{N_k}x^{l},
\end{gather}
 where the symbol $[\ ]$ means greatest integer part. Also the functions
$\eta_{N_{n}}^{\alpha_{n}}\!\left(\frac{1}{\beta}\right)$,
$\!{\eta_{N_{n-1},N_{n}}^{\alpha_{n-1},\alpha_{n}}}\!\left(\frac{1}{\beta}\right)$,
$\ldots$ and
${\eta_{N_{2},N_{3},\ldots,N_{n}}^{\alpha_{2},\alpha_{3},\ldots,\alpha_{n}}}\!
\left(\frac{1}{\beta}\right)$ are defined in the following forms:
\begin{gather*}
 \eta_{N_{n}}^{\alpha_{n}}\left(\frac{1}{\beta}\right)=
\beta\left(\frac{\alpha_nA_{N_{n}}\left(\frac{1}{\beta}\right)}
{B_{N_{n}}\left(\frac{1}{\beta}\right)}\right)^2\!,\quad
\eta_{N_{n-1},N_{n}}^{\alpha_{n-1},\alpha_{n}}\left(\frac{1}{\beta}\right)=
\beta\left(\frac{\alpha_{n-1}A_{N_{n-1}}\Big(\mbox{\raisebox{1.5mm}{$\frac{1}{\eta_{N_{n}}^
{\alpha_{n}}\left(\frac{1}{\beta}\right)}$}}\Big)}
{B_{N_{n-1}}\Big(\mbox{\raisebox{1.5mm}{$\frac{1}{\eta_{N_{n}}^{\alpha_{n}}
\left(\frac{1}{\beta}\right)}$}} \Big)}\right)^2\!,  \\
 \ldots\ldots\ldots\ldots\ldots\ldots\ldots\ldots\ldots\ldots\ldots\ldots\ldots\ldots
 \ldots\ldots\ldots\ldots \\
\eta_{N_{2},N_{3},\ldots,N_{n}}^{\alpha_{2},\alpha_{3},\ldots,\alpha_{n}}
\left(\frac{1}{\beta}\right)= \beta\left(\frac{\alpha_2A_{N_{2}}
\Big(\mbox{\raisebox{1.5mm}{$\frac{1}{\eta_{N_{3},N_{4},\ldots,N_{n}}^
{\alpha_{3},\alpha_{4},\ldots,\alpha_{n}}\left(\frac{1}{\beta}\right)}$}}\Big)}
{B_{N_{2}}
\Big(\mbox{\raisebox{1.5mm}{$\frac{1}{\eta_{N_{3},N_{4},\ldots,N_{n}}^
{\alpha_{3},\alpha_{4},\ldots,\alpha_{n}}\left(\frac{1}{\beta}\right)}$}}\Big)}\right)^2.
\end{gather*}
 As we see the above measure is defined
only for $\beta > 0$, hence from the
 relations (3.4), it follows that these maps  are chaotic in the region of
 the parameters space which lead to positive solution of
$\beta$. Taking the limits of $\beta\longrightarrow 0_+$ and
$\beta\longrightarrow \infty$ in the relation $(3.4)$ respectively
one can show that the chaotic regions is:
$\prod\limits_{k=1}^{n}\frac{1}{N_k}<
\prod\limits_{k=1}^{n}\alpha_k <\prod\limits_{k=1}^{n} N_k $ for
odd integer values of $N_1,N_2,\ldots,N_n$ and if one of the
integers happens to become even, then the chaotic region in the
parameter space is defined by $\alpha_k>0$, for $k=1,2,\ldots,n$
and
 $\prod\limits_{k=1}^{n}\alpha_k <\prod\limits_{k=1}^{n} N_k $ if one of the
integers happens to become even, respectively. Out of these
regions they have only period one stable fixed points.

 In order to
prove that measure (3.3) satisfies equation (3.1), with $\beta $
given by relation~(3.4), it is rather convenient to consider the
conjugate map
$\tilde{\Phi}_{N_{1},N_{2},\ldots,N_{n}}^{\alpha_1,\alpha_2,\ldots,\alpha_n}(x),
$ with measure $\tilde{\mu}_{\tilde{\Phi}_{N_1,N_2\ldots,N_n}^
{{\alpha_1,\alpha_2\ldots,\alpha_n}}}$ denoted by
$\tilde{\mu}_{\tilde{\Phi}}$ related to the measure
$\mu_{\Phi_{N_1,N_2\ldots,N_n}^{{\alpha_1,\alpha_2\ldots,\alpha_n}}}$
denoted by $\mu_{\Phi}$ through the following relation:
\[
\tilde{\mu_{\tilde{\Phi}}}(x)=\frac{1}{(1+x)^2}\mu_{\Phi}\left(\frac{1}{1+x}\right).
\]
Denoting
$\tilde{\Phi}_{N_{1},N_{2},\ldots,N_{n}}^{\alpha_1,\alpha_2,\ldots,\alpha_n}(x)$
 by $y$ and inverting it, we get:
 \begin{gather*}
x_{k_1}=\tan^2\left(\frac{1}{N_1}\arctan\sqrt{y\alpha_1^2}+\frac{k_1\pi}{N_1}\right),\\
x_{k_1,k_2}=\tan^2\left(\frac{1}{N_2}\arctan\sqrt{x_{k_1}\alpha_2^2}+\frac{k_2\pi}{N_2}\right),\\
\ldots\ldots\ldots\ldots\ldots\ldots\ldots\ldots\ldots\ldots\ldots\ldots\ldots\ldots\ldots
\ldots\ldots\ldots\\
x_{k_1,k_2,\ldots,k_n}=\tan^2\left(\frac{1}{N_n}\arctan\sqrt{x_{k_1,k_2,\ldots,k_{n-1}}
\alpha_n^2}+\frac{k_n\pi}{N_n}\right), \end{gather*} for
$k_{j}=1,\ldots,N_{j}$ and $j=1,\ldots,n$.

 Then by taking the derivative of $x_{k_1,k_2,\ldots,k_n}$ with respect to $y$, we obtain:
\begin{gather}
\hspace*{-27pt}\left|\frac{dx_{k1,k_2,\ldots,k_n}}{dy}\right|=
\left(\prod_{k=1}^{n}\frac{\alpha_k}{N_k}\right)
\sqrt{\frac{x_{k_1,k_2,\ldots,k_n}}{y}}\nonumber\\
\hspace*{-27pt}\times\frac{(1+x_{k_1,k_2,\ldots,k_n})(1+x_{k_2,k_3,\ldots,k_n})
\cdots(1+x_{k_{n-1},k_n})(1+x_{k_n})}
{(1+\alpha_n^2x_{k_2,k_3,\ldots,k_n})
(1+\alpha_{n-1}^2x_{k_3,k_4,\ldots,k_n})\cdots
(1+\alpha_3^2x_{k_{n-1},k_n}) (1+\alpha_2^2x_{k_n})
(1+\alpha_1^2y)}\end{gather} to be substituted in equation (3.1).
In derivation of above formula we have used chain rule property of
the derivative of composite functions.

 Substituting the above
result in equation (3.1), we have:
\begin{gather*}
\hspace*{-27pt}\tilde{\mu}_{\tilde{\Phi}}(y)\sqrt{y}\left(1+\alpha_1^2y\right)=
\left(\prod_{k=1}^{n}\frac{\alpha_k}{N_k}\right)\sum_{k_1}\sum_{k_2}
\cdots\sum_{k_n}\sqrt{x_{k_1,k_2,\ldots,k_n}}\\
\hspace*{-27pt}\times
\frac{(1+x_{k_1,k_2,\ldots,k_n})(1+x_{k_2,k_3,\ldots,k_n})
\cdots(1+x_{k_{n-1},k_n})(1+x_{k_n})}
 { (1+\alpha_n^2x_{k_2,k_3,\ldots,k_n})
 (1+\alpha_{n-1}^2x_{k_3,k_4,\ldots,k_n})\cdots
 (1+\alpha_3^2x_{k_{n-1},k_n})(1+\alpha_2^2x_{k_n})}
 \tilde{\mu}_{\tilde{\Phi}}(x_{k_1,k_2,\ldots,k_n}).
 \end{gather*}
  Now considering the following equation for the invariant
measure $\tilde{\mu}_{\tilde{\Phi}}(y)$:
\begin{equation}
\tilde{\mu}_{\tilde{\Phi}}(y)=\frac{1}{\sqrt{y}(1+\beta y)},
\end{equation}
then the FP-equation reduced to:
\begin{gather*}
\hspace*{-27pt}\frac{1+\alpha_1^2y}{1+\beta
y}=\left(\prod_{k=1}^{n}\frac{\alpha_k}{N_k}\right)\\
\hspace*{-27pt}\times\!\sum_{k_1=1}^{N_1}
\sum_{k_2=1}^{n_2}\!\cdots\!\!\sum_{k_n=1}^{N_n}\!
\left(\!\frac{(1+x_{k_1,k_2,\ldots,k_n}\!)(1+x_{k_2,k_3,\ldots,k_n}\!)
\cdots(1+x_{k_{n-1},k_n}\!)(1+x_{k_n}\!)}
{(1+\alpha_n^2x_{k_2,k_3,\ldots,k_n}\!)(1+\alpha_{n-1}^2
 x_{k_3,k_4,\ldots,k_n}\!)\cdots(1+\alpha_3^2x_{k_{n-1},k_n}\!)
 (1+\alpha_2^2x_{k_n}\!)}\!
\right)\!.\!
\end{gather*}
 Now using the follow identity:
 \begin{equation}
\frac{\alpha}{N}\sum_{k=0}^{N}\frac{1+\alpha^{2}x_{k}}{1+\beta
x_{k}}=\frac{1+\alpha^{2}y}{\left( \frac{B\left(
\frac{1}{\beta}\right)}{\alpha A\left(
\frac{1}{\beta}\right)}+\beta\left( \frac{\alpha A\left(
\frac{1}{\beta}\right)}{B\left(
\frac{1}{\beta}\right)}\right)y\right)},
\end{equation}
   for a one-parameter chaotic map $y=\Phi_{N}^{\alpha}(x)$ (Its
   proof is given in Appendix~B), we obtain:
\begin{gather*}
\frac{\alpha_n}{N_n}\sum_{k_n=1}^{N_{n}}\frac{1+x_{k_1,k_2,\ldots,k_n}}
{1+\beta x_{k_1,k_2,\ldots,k_n}}
=\frac{\alpha_nA_{N_{n}}\left(\frac{1}
{\beta}\right)}{B_{N_{n}}\left(\frac{1}{\beta}\right)}
\frac{1+\alpha_n^{2}x_{k_1,k_2,\ldots,k_{n-1}}}
{1+\eta_{N_{n}}^{\alpha_n}\left(\frac{1}{\beta}\right)
x_{k_1,k_2,\ldots,k_{n-1}}},\\
 \frac{\alpha_{n-1}\alpha_{n}}{N_{n-1}N_{n}}
\sum_{k_{n-1}=1}^{N_{n-1}}
\sum_{k_n=1}^{N_{n}}\frac{(1+x_{k_1,k_2,\ldots,k_{n-1}})
(1+x_{k_1,k_2,\ldots,k_{n}})}
{(1+\alpha_{n-1}x_{k_1,k_2,\ldots,k_{n-1}})(1+\beta
x_{k_1,k_2,\ldots,k_n})}\\ = \frac{\alpha_n\alpha_{n-1}
A_{N_{n}}\left(\frac{1}{\beta}\right)A_{N_{n-1}}
\Big(\mbox{\raisebox{1.5mm}{$\frac{1}
{\eta_{N_{n}}^{\alpha_n}\left(\frac{1}{\beta}\right)}$}}\Big)}
{B_{N_{n}}\left(\frac{1}{\beta}\right)B_{N_{n-1}}
\Big(\mbox{\raisebox{1.5mm}{$\frac{1}
{\eta_{N_{n}}^{\alpha_n}\left(\frac{1}{\beta}\right)}$}}\Big)}\times
\frac{1+\alpha_{n-1}^{2}x_{k_1,k_2,\ldots,k_{n-2}}} {1+
\eta_{N_{n-1},N_{n}}^{\alpha_{n-1},\alpha_n}\left(\frac{1}{\beta}\right)
x_{k_1,k_2,\ldots,k_{n-2}}},\\
   \ldots\ldots\ldots\ldots\ldots\ldots\ldots\ldots\ldots\ldots\ldots\ldots\ldots\ldots\ldots\ldots\ldots\ldots\ldots\ldots\ldots\ldots\ldots\ldots\ldots\\
\left(\prod_{k=1}^{n}\frac{\alpha_k}{N_k}\right)\sum_{k_1=1}^{N_1}
\sum_{k_2=1}^{n_2}\cdots\\
 \times\sum_{k_n=1}^{N_n}
\left(\frac{(1+x_{k_1,k_2,\ldots,k_n})(1+x_{k_2,k_3,\ldots,k_n})
\cdots(1+x_{k_{n-1},k_n})(1+x_{k_n})
 }{ (1+\alpha_n^2x_{k_2,k_3,\ldots,k_n})
 (1+\alpha_{n-1}^2x_{k_3,k_4,\ldots,k_n})
 \cdots(1+\alpha_3^2x_{k_{n-1},k_n})(1+\alpha_2^2x_k{_{n}})}
\right)\\ = \prod _{k=1}^{n}\alpha_k
\times\frac{A_{N_{n}}\left(\frac{1}{\beta}\right)}
{B_{N_{n}}\left(\frac{1}{\beta}\right)}\times
\frac{A_{N_{n-1}}\Big(\mbox{\raisebox{1.5mm}{$\frac{1}{\eta_{N_{n}}^{\alpha_{n}}
\left(\frac{1}{\beta}\right)}$}}\Big)}
{B_{N_{n-1}}\Big(\mbox{\raisebox{1.5mm}{$\frac{1}{\eta_{N_{n}}^{\alpha_{n}}
\left(\frac{1}{\beta}\right)}$}}\Big)}\\
 \times \frac{A_{N_{n-2}}\Big(\mbox{\raisebox{1.5mm}{$\frac{1}{\eta_{N_{n-1},N_{n}}^
{\alpha_{n-1},\alpha_{n}}\left(\frac{1}{\beta}\right)}$}}\Big)}
{B_{N_{n-2}}\Big(\mbox{\raisebox{1.5mm}{$\frac{1}{\eta_{N_{n-1},N_{n}}^
{\alpha_{n-1},\alpha_{n}}\left(\frac{1}{\beta}\right)}$}}\Big)}
\times \cdot \times
\frac{A_{N_{1}}\Big(\mbox{\raisebox{1.5mm}{$\frac{1}{\eta_{N_{2},N_{3},\ldots,N_{n}}
^{\alpha_{2},\alpha_{3},\ldots,a_{n}}\left(\frac{1}{\beta}\right)}$}}\Big)}
{B_{N_{1}}\Big(\mbox{\raisebox{1.5mm}{$\frac{1}{\eta_{N_{2},N_{3},\ldots,N_{n}}
^{\alpha_{2},\alpha_{3},\ldots,\alpha_{n}}\left(\frac{1}{\beta}\right)}$}}\Big)}
\frac{1+\alpha_1^2y}{1+\eta_{N_{1},N_{2},\ldots,N_{n}}
^{\alpha_{1},\alpha_{2},\ldots,\alpha_{n}}\left(\frac{1}{\beta}\right)y}.
\end{gather*}
Now by inserting the right side of last relation  in (3.5), we
get: \begin{gather*}
 \frac{1+\alpha_1^{2}y}{1+\beta y}=\prod
_{k=1}^{n}\alpha_k
\frac{A_{N_{n}}\left(\frac{1}{\beta}\right)}{B_{N_{n}}\left(\frac{1}{\beta}\right)}\times
\frac{A_{N_{n-1}}
\Big(\mbox{\raisebox{1.5mm}{$\frac{1}{\eta_{N_{n}}^{\alpha_{n}}\left(\frac{1}{\beta}\right)}
$}}\Big)}
{B_{N_{n-1}}\Big(\mbox{\raisebox{1.5mm}{$\frac{1}{\eta_{N_{n}}^{\alpha_{n}}
\left(\frac{1}{\beta}\right)}$}}\Big)}\\ \qquad {}\times
\frac{A_{N_{n-2}}\Big(\mbox{\raisebox{1.5mm}{$\frac{1}{\eta_{N_{n-1},N_{n}}^
{\alpha_{n-1},\alpha_{n}}\left(\frac{1}{\beta}\right)}$}}\Big)}
{B_{N_{n-2}}\Big(\mbox{\raisebox{1.5mm}{$\frac{1}{\eta_{N_{n-1},N_{n}}^
{\alpha_{n-1},\alpha_{n}}\left(\frac{1}{\beta}\right)}$}}\Big)}
\times \cdot \times
\frac{A_{N_{1}}\Big(\mbox{\raisebox{1.5mm}{$\frac{1}{\eta_{N_{2},N_{3},\ldots,N_{n}}^
{\alpha_{2},\alpha_{3},\ldots,a_{n}}\left(\frac{1}{\beta}\right)}$}}\Big)}
{B_{N_{1}}\Big(\mbox{\raisebox{1.5mm}{$\frac{1}{\eta_{N_{2},N_{3},\ldots,N_{n}}
^{\alpha_{2},\alpha_{3},\ldots,\alpha_{n}}\left(\frac{1}{\beta}\right)}$}}\Big)}
\frac{1+\alpha_1^2y}{1+\eta_{N_{1},N_{2},\ldots,N_{n}}^
{\alpha_{1},\alpha_{2},\ldots,\alpha_{n}}\left(\frac{1}{\beta}\right)y}.
\end{gather*}
 We see that the above relation holds true provided that the parameter
 $\beta$ fulfills the relation~(3.4).

\section{Kolmogrov--Sinai entropy}
KS-entropy or metric entropy \cite{sinai, Dorf2} measures how
chaotic a dynamical system is and it is proportional to the rate
at which information about the state of dynamical system is lost
in the course of time or iteration. Therefore, it can also be
defined as the average rate of information loss for a discrete
measurable dynamical system
$(\Phi_{N_1,N_2,\ldots,N_n}^{\alpha_1,\alpha_2,\ldots,\alpha_n}(x),\mu)$.
By introducing a partition $\alpha={A_c} (n_1,\ldots,n_{\gamma})$
of the interval $[0,1]$ into individual laps $A_i$, one can define
the usual entropy associated with the partition by:
\[
H(\mu,\gamma)=-\sum^{n(\gamma)}_{i=1}m(A_c)\ln{m(A_c)}, \]
 where
$m(A_c)=\int_{n\in{A_i}}\mu(x)dx$ is the invariant measure of $
A_i$. By defining $n$-th refining $\gamma(n)$ of  $ \gamma$ as:
\[
\gamma^{n}=\bigcup^{n-1}_{k=0}
\left(\Phi_{N_1,N_2,\ldots,N_n}^{\alpha_1,\alpha_2,\ldots,\alpha_n}(x)\right)^{-(k)}(\gamma),
\]
then entropy per unit step of refining is defined by:
\[
h\left(\mu,\Phi_{N_1,N_2,\ldots,N_n}^{\alpha_1,\alpha_2,\ldots,\alpha_n}(x)
,\gamma\right)=\lim_{n\rightarrow
\infty}\left(\frac{1}{n}H(\mu,\gamma)\right). \]
 Now, if the size
of individual laps of $\gamma(N)$ tends to zero as n increases,
then the above entropy is known as KS-entropy, that is:
\[
h\left(\mu,\Phi_{N_1,N_2,\ldots,N_n}^{\alpha_1,\alpha_2,\ldots,\alpha_n}(x)\right)
=h\left(\mu,\Phi_{N_1,N_2,\ldots,N_n}^{\alpha_1,\alpha_2,\ldots,\alpha_n}(x),
\gamma\right). \]
 KS-entropy, which is a quantitative measure of
the rate of information loss with the refining, may also be
written as:
\begin{equation}
h\left(\mu,\Phi_{N_1,N_2,\ldots,N_n}^{\alpha_1,\alpha_2,\ldots,\alpha_n}(x)\right)
=\int \mu(x)dx \ln
\left|\frac{d}{dx}\Phi_{N_1,N_2,\ldots,N_n}^{\alpha_1,\alpha_2,\ldots,\alpha_n}
(x)\right|,
\end{equation}
which is also a statistical mechanical expression for the Lyapunov
characteristic exponent, that is the mean divergence rate of two
nearby orbits. The measurable dynamical system
$(\Phi_{N_1,N_2,\ldots,N_n}^{\alpha_1,\alpha_2,\ldots,\alpha_n}(x),\mu)$
is chaotic for $h>0$ and predictive for $h=0$.

In order to calculate the KS-entropy of the maps
$\Phi_{N_1,N_2,\ldots,N_n}^{\alpha_1,\alpha_2,\ldots,\alpha_n}(x)$,
it is rather convenient to consider their conjugate maps given by
$(2.12)$, since it can be shown that KS-entropy is a kind of
topological invariant, that is, it is preserved under conjugacy
map. Hence, we have:
\[
h\left(\mu,\Phi_{N_1,N_2,\ldots,N_n}^{\alpha_1,\alpha_2,\ldots,\alpha_n}(x)\right)
=h\left(\tilde{\mu},\tilde{\Phi}_{N_1,N_2,\ldots,N_n}^{\alpha_1,
\alpha_2,\ldots,\alpha_n}(x)\right). \]
 Using the integral
(4.1), the KS-entropy of
$\Phi_{N_1,N_2,\ldots,N_n}^{\alpha_1,\alpha_2,\ldots,\alpha_n}(x)$
can be written as:
\begin{gather*}
 h\left(\mu,\Phi_{N_1,N_2,\ldots,N_n}^{\alpha_1,\alpha_2,\ldots,\alpha_n}(x)\right)\\
=
\frac{1}{\phi}\int_{0}^{\infty}\frac{\sqrt{\beta}dx}{\sqrt{x}(1+\beta
x)}\ln\left|\frac{d}{dy_{N_2,N_3,\cdot,N_n}}\left(\frac{1}{\alpha_1^{2}}\tan^{2}
\left(N_1 \arctan\sqrt{y_{N_2,N_3,\cdot,N_n}}\right)
\right.\right.\\ \left.\left. \times\frac{d}{
dy_{N_3,N_4,\cdot,N_n}}\left(\frac{1}{\alpha_2^{2}}\tan^{2}
\left(N_2 \arctan\sqrt{y_{N_3,N_4,\cdot,N_n}}\right)\cdots
\frac{d}{dx}\frac{1}{\alpha_n^{2}}\tan^{2}(N_n
\arctan\sqrt{x})\right)\right)\right|
\end{gather*}
 or
 \begin{gather}
h\left(\mu,\Phi_{N_1,N_2,\ldots,N_n}^{\alpha_1,\alpha_2,\ldots,\alpha_n}(x)\right)\nonumber\\
\qquad{}=\frac{1}{\pi}\int_{0}^{\infty}\frac{\sqrt{\beta}dx}{\sqrt{x}(1+\beta
x)}\ln\left|\frac{d}{dy_{N_2,N_3,\cdot,N_n}}\left(\frac{1}{\alpha_1^{2}}\tan^{2}\left(N_1
\arctan\sqrt{y_{N_2,N_3,\cdot,N_n}}\right)\right)\right|\nonumber\\
\qquad{} +
\frac{1}{\pi}\int_{0}^{\infty}\frac{\sqrt{\beta}dx}{\sqrt{x}(1+\beta
x)}\ln\left|\frac{d}{dy_{N_3,N_4,\cdot,N_n}}\left(\frac{1}{\alpha_2^{2}}\tan^{2}\left(N_2
\arctan\sqrt{y_{N_3,N_4,\cdot,N_n}}\right)\right)\right|\nonumber\\
\qquad{}+ \cdots +
\frac{1}{\pi}\int_{0}^{\infty}\frac{\sqrt{\beta}dx}{\sqrt{x}(1+\beta
x)}\ln\left|\frac{d}{dy_{N_n}}\left(\frac{1}{\alpha_{n-1}^{2}}\tan^{2}\left(N_{n-1}
\arctan\sqrt{y_{N_n}}\right)\right)\right|\nonumber\\ \qquad{}+
\frac{1}{\pi}\int_{0}^{\infty}\frac{\sqrt{\beta}dx}{\sqrt{x}(1+\beta
x)}\ln\left|\frac{d}{dx}\left(\frac{1}{\alpha_n^{2}}\tan^{2}\left(N_n
\arctan\left(\sqrt{x}\,\right)\right)\right)\right|, \end{gather}
where
\begin{gather} y_{N_n}=\frac{1}{\alpha_n^{2}}\tan^{2}\left(N_n
\arctan(\sqrt{x})\right),
\\
y_{N_{n-1},N_n}=\frac{1}{\alpha_{n-1}}\tan^{2}\left(N_{n-1}\arctan(\sqrt{y_{N_n}}))\right),
\\
    \ldots\ldots\ldots\ldots\ldots\ldots \ldots\ldots\ldots\ldots\ldots\ldots\ldots\ldots\ldots\nonumber\\
y_{N_2,N_3,\cdot,N_n}=\frac{1}{\alpha_1^{2}}\tan^{2}\left(N_1
\arctan(\sqrt{y_{N_3,N_4,\cdot,N_n}})\right).
\end{gather}
 Now, we calculate the integrals appearing above in the expression for the
entropy separately. The last integral is calculated in appendix~C
which reads:
\begin{gather}
\frac{1}{\pi}\int_{0}^{\infty}\!\frac{\sqrt{\beta}dx}{\sqrt{x}(1+\beta
x)}\ln\left|\frac{d}{dx}\left(\frac{1}{\alpha_n^{2}}\tan^{2}\left(N_n
\arctan\left(\sqrt{x}\,\right)\right)\right)\right|\nonumber\\
\qquad{}
=\ln\left(\frac{N_n\left(1+\beta+2\sqrt{\beta}\,\right)^{N_n-1}}{A_{N_{n}}
\left(\frac{1}{\beta}\right)B_{N_{n}}\left(\frac{1}{\beta}\right)}\right).
\end{gather}
In order to calculate the integral before the last one in $(4.2)$,
that is the following integral,
\begin{equation}
\frac{1}{\pi}\int_{0}^{\infty}\frac{\sqrt{\beta}dx}{\sqrt{x}(1+\beta
x)}\ln\left|\frac{d}{dy_{N_n}}\left(\frac{1}{\alpha_{n-1}^{2}}\tan^{2}\left(N_{n-1}
\arctan\sqrt{y_{N_n}}\right)\right)\right|,
\end{equation} first we make the
following change of variable by inverting the relation (4.3)
\[
x_{k_{n-1}}=\tan^2\left(\frac{1}{N_{n-1}}\arctan
\left(\sqrt{y_{N_{n}}}\alpha_{n-1}^2\right)+\frac{k_{n-1}
\pi}{N_{n-1}}\right),\qquad k_{n-1}=1,\ldots,N_{n-1}. \]
  Then the
integral (4.7) is reduced to:
\[
\sum_{k_{n-1}=1}^{N_{n-1}}\frac{1}{\pi}\int_{x_{k_{n-1}}^{i}}^
{x_{k_{n-1}}^{f}}\frac{\sqrt{\beta}dx_{k_{n-1}}}{\sqrt{x_{k_{n-1}}}
(1+\beta x_{k_{n-1}})}\ln\left| \frac{d}{dy_{N_n}}
\left(\frac{1}{\alpha_{n-1}^{2}}\tan^{2}\left(N_{n-1}\arctan\sqrt{y_{N_n}}\right)\right)\right|,
\]
 where $x_{k_{n-1}}^{i}$ and $x_{k_{n-1}}^{f}$  ($
k_{n-1}=1,2,\ldots,N_{n-1}$) denote the initial and end points of
$k$-th branch of the inversion of function
$y_{N_n}=\left(\frac{1}{\alpha_n^{2}}\tan^{2}\left(N_n
\arctan\sqrt{x}\right)\right)$ respectively. Now, by inserting the
derivative of $x_{k_{n-1}}$ with respect to $ y_{{N_n}}$ in the
above relation and changing the order of sum and integration, we
get:
\begin{gather*}
 \frac{1}{\pi}\int_{0}^{\infty}\sum_{k_{n-1}=1}^{N_{n-1}}
\sqrt{\beta}dy_{N_{n}}\frac{\alpha_{n-1}\sqrt{x_{k_{n-1}}}
(1+x_{k_{n-1}})}
{N_{n-1}\sqrt{y_{n_{n}}}\left(1+\alpha_{n-1}^2y_{N_{n}}\right)
{\sqrt{x_{k_{n-1}}}}(1+\beta x_{k_{n-1}})}\\
\qquad {}\times\ln\left|\frac{d}
{dy_{N_n}}\left(\frac{1}{\alpha_{n-1}^{2}}\tan^{2}(N_{n-1}\arctan\sqrt{y_{N_n}})\right)\right|.
\end{gather*}
By using the formula (B.6) of Appendix~B, we get:
\begin{gather*}
\frac{1}{\pi}\int_{0}^{\infty}
\frac{\sqrt{\beta}dy_{n}}{\sqrt{y_{n}}}
\left(\frac{B_{N_{n}}\left(\frac{1}{\beta}\right)}{\alpha_nA_{N_{n}}\left(\frac{1}{\beta}\right)}+
\beta\frac{\alpha_nA_{N_{n}}\left(\frac{1}{\beta}\right)}
{B_{N_{n}}\left(\frac{1}{\beta}\right)}y_{N_{n}}\right)\\
\qquad {}\times \ln\left|\frac{d}{dy_{N_n}}
\left(\frac{1}{\alpha_{n-1}^{2}}\tan^{2}\left(N_{n-1}\arctan\sqrt{y_{N_n}}\right)\right)\right|.
\end{gather*}
Finally, through calculating the above integral with the
prescription of Appendix~C, we obtain:
\[
\ln\left(\frac{N_{n-1}\left(1+\eta_{N_{n}}^{\alpha_{n}}+2\sqrt{\eta_{N_{n}}^{\alpha_{n}}}
\right)^{N_{n-1}-1}}
{A_{N_{n-1}}\left(\eta_{N_{n}}^{\alpha_{n}}\right)
B_{N_{n-1}}\left(\eta_{N_{n}}^{\alpha_{n}}\right)}\right).
\]
 Similarly, we
can calculate the other integrals appearing in the expression for
the entropy of the composed maps given in (4.2):
\begin{gather*}
h\left(\mu,\Phi_{N_1,N_2,\ldots,N_n}^{\alpha_1,\alpha_2,\ldots,\alpha_n}(x)\right)\\
=\frac{1}{\pi}\int_{0}^{\infty}\!\frac{\sqrt{\beta}dx}{\sqrt{x}(1+\beta
x)}\ln\left|\frac{d}{dy_{N_k,N_{k+1},\cdot,N_n}}\left(\frac{1}{\alpha_{k-1}^{2}}\tan^{2}\left(N_{k-1}
\arctan\sqrt{y_{N_k,N_{k+1},\cdot,N_n}}\right)\right)\right|\\
=\ln\left(\frac{N_{k-1}\left(1+ \eta_{N_{k},N_{k-1},\ldots,N_{n}}^
{\alpha_{k},\alpha_{k+1},\ldots,\alpha_{n}}\left(\frac{1}{\beta}\right)+
2\sqrt{\eta_{N_{k},N_{k-1},\ldots,N_{n}}^
{\alpha_{k},\alpha_{k+1},\ldots,\alpha_{n}}\left(\frac{1}{\beta}\right)}\right)^{(N_{k-1})-1}}
{A_{N_{k-1}}\left(\eta_{N_{k},N_{k-1},\ldots,N_{n}}^
{\alpha_{k},\alpha_{k+1},\ldots,\alpha_{n}}\left(\frac{1}{\beta}\right)\right)
B_{N_{k-1}}\left(\eta_{N_{k},N_{k-1},\ldots,N_{n}}^
{\alpha_{k},\alpha_{k+1},\ldots,\alpha_{n}}\right)\left(\frac{1}{\beta}\right)}\right)
\end{gather*}
for $k=1,2,\ldots,n$.

Finally, summing the
above integral, we get the following expression for the entropy of
these maps:
\begin{gather}
h\left(\mu,\Phi_{N_1,N_2,\ldots,N_n}^{\alpha_1,\alpha_2,\ldots,\alpha_n}(x)\right)=
 \ln\Bigg\{ \Bigg[(N_1N_2\cdots
N_n)\left(1+\sqrt{\beta}\right)^{2(N_{n}-1)}
\nonumber\\
\times \left(1+\sqrt{\eta_{N_{n}}^{\alpha_{n}}\left(\frac{1}{\beta}\right)}\right)^
{2(N_{n-1}-1)}\!\!\cdots\left(1+\sqrt{\eta_{N_{2},N_{3},\ldots,N_{n}}^
{\alpha_{2},\alpha_{3},\ldots,\alpha_{n}}\left(\frac{1}{\beta}\right)}\right)^{2(N_{1}-1)}\Bigg]
\Bigg/ \Bigg[A_{N_n}(\beta)\nonumber\\
\times B_{N_n}(\beta)A_{N_{n-1}}
\left(\eta_{N_{n}}^{\alpha_{n}}\left(\frac{1}{\beta}\right)\right)
B_{N_{n-1}}\left(\eta_{N_{n}}^{\alpha_{n}}\left(\frac{1}{\beta}\right)\right) \cdots\nonumber\\
\times
A_{N_{1}}\left(\eta_{N_{2},N_{3},\ldots,N_{n}}^
{\alpha_{2},\alpha_{3},\ldots,\alpha_{n}}\left(\frac{1}{\beta}\right)\right)
B_{N_{1}}\left(\eta_{N_{2},N_{3},\ldots,N_{n}}^
{\alpha_{2},\alpha_{3},\ldots,\alpha_{n}}\left(\frac{1}{\beta}\right)\right)\Bigg]\Bigg\}.
\end{gather}
Using the formulas (4.8), one can show that KS-entropy of one
parameter families has the following asymptotic behavior:
\begin{gather*}
h\left(\mu,\Phi_{N}\left(x,\alpha=N+0^{-}\right)\right)\sim
(N-\alpha)^{\frac{1}{2}},\\
h\left(\mu,\Phi_{N}\left(x,\alpha=\frac{1}{N}+0^{+}\right)\right)\sim
\left(\alpha-\frac{1}{N}\right)^{\frac{1}{2}},
\end{gather*}
near the bifurcation points, that is $\beta \longrightarrow 0$ as
$ \beta \longrightarrow \infty $. The above asymptotic behaviors
indicate that one-parameter maps $\Phi_{N}^{\alpha}(x)$ belong to
the same universality class which are different from the
universality class of pitch fork bifurcating maps, but their
asymptotic behavior is similar to class of intermittent
 maps~\cite{pomeau}. Even though, intermittency can not occur in these
maps for any values of parameter $\alpha$, the maps
$\Phi_{N}^{\alpha}(x)$ and their n-composition $\Phi^{(n)}$ do not
have minimum values other than zero and maximum values other
than~$1$ in the interval $[0,1]$.

 Also by imposing the relations between the parameters
$\alpha_k$, $k=1,2,\ldots,n$ which are consistent with the relation
(3.4), we can reduce these maps to other many-parameter families
of maps with the number of the parameters less than~$n$.
Particularly by imposing enough relations, we can reduce them to
one-parameter families of chaotic maps with an arbitrary
asymptotic behavior as the parameter takes the limiting values.
Hence we can construct chaotic maps with arbitrary universality
class. As an illustration, we consider the chaotic map
$\Phi_{2,2}^{\alpha_1,\alpha_2}(x)$. Using the formula (4.8),
we have:
\[
h\left(\mu,\Phi_{2,2}^{\alpha_1,\alpha_2}(x)\right)=\ln\frac{(1+\sqrt{\beta})^2
\left(2\sqrt{\beta}+\alpha_2(1+\beta)\right)^2}{(1+\beta)
\left(4\beta+\alpha_2^2(1+\beta)^2\right)}
\]
 with the following relation
among the parameters $\alpha_1,\alpha_2$ and $\beta$:
\[
\alpha_1\left(4\beta+\alpha_2^2(1+\beta)^2\right)=4\alpha_2\beta(1+\beta)
\]
which is obtained from the relation (3.4). Now choosing
$\beta=\alpha_2^{\nu}$, $0<\nu<2$, the above relation reduces to:
\[
\alpha_1=\frac{4\alpha_2^{1+\nu}(1+\alpha_2^{\nu})}
{\alpha_2^2(1+\alpha_2^{\nu})^2+4\alpha_2^{\nu}}
\]
 and entropy
given by (4.8) reads:
\[
h\left(\mu,\Phi_{2,2}^{\alpha_2}(x)\right)=\ln\frac{\left(1+\alpha_2^{\frac{\nu}{2}}\right)^2
\left(2\alpha_2^{\frac{\nu}{2}}+\alpha_2(1+\alpha_2^{\nu})\right)^2}
{(1+\alpha_2^{\nu})\left(4\alpha_2^{\nu}+\alpha_2^2(1+\alpha_2^{\nu})^2\right)}
\]
 which has the following asymptotic behavior near
$\alpha_2\longrightarrow 0$ and $\alpha_2\longrightarrow\infty$:
\begin{gather*}
h\left(\mu,\Phi_{2,2}^{\alpha_2}(x)\right)\sim\alpha_2^{\frac{\nu}{2}}\qquad
\mbox{as}\quad \alpha_2\longrightarrow 0, \\
h\left(\mu,\Phi_{2,2}^{\alpha_2}(x)\right)\sim\left(\frac{1}{\alpha_2}\right)^
{\frac{\nu}{2}}\qquad\mbox{as}\quad
\alpha_2\longrightarrow\infty.
\end{gather*}
 The above asymptotic behaviours indicate that for an  arbitrary value
of $0<\nu<2$ the maps $\Phi_{2,2}^{\alpha_2}(x)$ belong to the
universality class which is different from the universality class
of one-parameter chaotic maps of $\Phi_{N}(x)$ (2.1) or the
universality class of pitch fork bifurcating maps.

\section{Conclusion}
 We have given hierarchy of exactly solvable many-parameter families of
  one-dimensional chaotic maps
with an invariant measure, that is measurable dynamical system
with an interesting property of being either chaotic or having
stable fixed point, and they bifurcate from a stable single
periodic state to chaotic one and vice-versa without having usual
period doubling or period-$n$-tupling scenario.

Again  this interesting property is due to the existence of
invariant measure for a region  of the parameters space of these
maps. Hence, to support this conjecture, it would be interesting
to find the other measurable one-parameter maps, specially coupled
or higher dimensional maps, which are under investigation.

\appendix

\section*{Appendix}
 \section{Shwartzian derivative}

The Shwarzian derivative ${S\Phi_N(x)}$ \cite{dev} is defined as:
\[
S\left(\Phi_N(x)\right)=\frac{\Phi_{N}'''(x)}{\Phi_{N}'(x)}-
\frac{3}{2}\left({\frac{\Phi_{N}''(x)}{\Phi_{N}'(x)}}\right)^2=
\left(\frac{\Phi_{N}''(x)}{\Phi_{N}'(x)}\right)'
-\frac{1}{2}\left(\frac{\Phi_{N}''(x)}{\Phi_{N}'(x)}\right)^2,
\]
 with a prime denoting a single differential. One can show that:
\[
S\left(\Phi_{N}(x)\right)=S\left(_2F_1\left(-N,N,\frac{1}{2},x\right)\right)\leq0,
\]
 since $\frac{d}{dx}\left(_2F_1\left(-N,N,\frac{1}{2},x\right)\right)$ can be written
as:
\[
\frac{d}{dx}\left(_2F_1\left(-N,N,\frac{1}{2},x\right)\right)=A\prod^{N-1}_{i=1}(x-x_i)
\]
 with $0 \leq{x_1}<{x_2}<{x_3}<\cdots <x_{N-1}\leq{1},$ then we
have:
\[
S\left(_2F_1\left(-N,N,\frac{1}{2},x\right)\right)
=\frac{-1}{2}\sum^{N-1}_{J=1}\frac{1}{(x-x_j)^2}-\left(\sum^{N-1}_{J=1}
\frac{1}{(x-x_j)}\right)^2<{0}.
\]
 Also, one can show the shwartzian derivative of composition of
function with negative shwartzian derivatives is negative too.

\section{Derivation of the formula (3.9)}
 In order to drive formula (3.9), we write the summation in its left
 side as:
 \begin{equation}
\frac{\alpha}{N}\sum_{k=0}^{N}\frac{1+\alpha^{2}x_{k}}{1+\beta
x_{k}}=\frac{\alpha}{\beta}+\left(\frac{\beta-1}{\beta^{2}}\right)
\frac{\partial}{\partial\beta^{-1}}\left(\ln\left(\prod_{k=1}^{N}(\beta^{-1}+x_{k})\right)\right).
\end{equation}
To evaluate the second term in the right side of (B.1), we
denote $\tilde{\Phi}_{N}^{\alpha}(x)$ by y therefore, the map
$\tilde{\Phi}_{N}^{(1)}(x,\alpha)=\frac{1}{\alpha^2}\tan^{2}(N\arctan\sqrt{x})$
can be written as:
 \begin{equation}
0=\alpha^{2}y\cos^{2}(N\arctan\sqrt{x})-\sin^{2}(N\arctan\sqrt{x}).
\end{equation}
 Now, we can write the equation (B.1) in the following form:
\begin{gather*}
\frac{\alpha}{N}\sum_{k=0}^{N}\frac{1+\alpha^{2}x_{k}}{1+\beta
x_{k}}={\frac{(-1)^{N}}{(1+x)^{N}}}\left(\alpha^{2}y\left(\sum_{k=0}
^{\left[\frac{N}{2}\right]}C_{2k}^{N}(-1)^{N}x^{k}\right)^{2}\right.\\
\qquad\left.{}-x\left(\sum_{k=0}
^{\left[\frac{N-1}{2}\right]}C_{2k+1}^{N}(-1)^{N}x^{k}\right)^{2}\right)
=\frac{\mbox{constant}}{(1+x)^{N}}\prod_{k=1}^{N}(x-x_{k}),
\end{gather*}
 where $x_{k}$ are the roots of equation (B.2) and they are given
 by
 \begin{equation}
x_k=\tan^2\left(\frac{1}{N}\arctan\sqrt{y\alpha^2}+\frac{k\pi}{N}\right),\qquad
k=1,\ldots,N.
\end{equation}
 Therefore, we have:
\begin{gather}
\frac{\partial}{\partial\beta^{-1}}\ln\left(\prod_{k=1}^{N}(\beta^{-1}+x_{k})\right)\nonumber\\
=\frac{\partial}{\partial\beta^{-1}}\ln\left[\left(1-\beta^{-1}\right)^{N}\left(\alpha^{2}y\cos^{2}
 \left(N\arctan\sqrt{-\beta^{-1}}\right)-\sin^{2}\left(N\arctan\sqrt{-\beta^{-1}}\right)\right)\right]\nonumber\\
=-\frac{N\beta}{\beta-1}+\frac{\beta
N\left(1+\alpha^2y\right)A\left(\frac{1}{\beta}\right)
}{\left(A\left(\frac{1}{\beta}\right)\right)^{2}\beta^{2}y+\left(B\left( \frac{1}{\beta}\right)\right)^2},
\end{gather}
with polynomials $ A(x)$ and $B(x)$ given in (3.5) and (3.6)
where in derivation of above formulas we have used the following
identities:
\begin{gather}
\cos\left(N\arctan\sqrt{x}\right)=\frac{A(-x)}{(1+x)^{\frac{N}{2}}},
\qquad
\sin\left(N\arctan\sqrt{x}\right)=\sqrt{x}\frac{B(-x)}{(1+x)^{\frac{N}{2}}}.
\end{gather}
By inserting the results (B.4) in (B.1), we get:
\begin{equation}
\frac{\alpha}{N}\sum_{k=0}^{N}\frac{1+\alpha^{2}x_{k}}{1+\beta
x_{k}}=\frac{1+\alpha^{2}y}{\left( \frac{B\left(
\frac{1}{\beta}\right)}{\alpha A\left( \frac{1}{\beta}\right)}+\beta\left( \frac{\alpha
A\left( \frac{1}{\beta}\right)}{B\left( \frac{1}{\beta}\right)}\right)y\right)}.
\end{equation}

\section{Derivation of the last integral of (4.2)}

Using the relations (3.5) and (3.6) the last integral
  of (4.2) can be written as:
    \begin{equation}
h(\mu,\Phi_{N}^{\alpha}(x))
=\frac{1}{\pi}\int_{0}^{\infty}\frac{\sqrt{\beta}dx}{\sqrt{x}(1+\beta
x)}\ln\left(\frac{N}{\alpha^{2}}\left|\frac{(1+x)^{N-1}B(-x)}{(A(-x))^{3}
}\right|\right).
\end{equation}
 We see that polynomials appearing in the numerator (denominator) of
integrand appearing on the right side of equation (C.1) have
$\frac{[N-1]}{2}$ $\left(\frac{[N]}{2}\right)$ simple roots denoted by
 $ x_{k}^{B}$, $k=1,\ldots,\left[\frac{N-1}{2}\right]$
$\left(x_{k}^{A}, \ k=1,\ldots,\left[\frac{N}{2}\right]\right)$
 in the interval $[0,\infty)$.
 Hence, we can write the above formula in the following form:
\[
h\left(\mu,\Phi_{N}^{(\alpha)}(x)\right)
=\frac{1}{\pi}\int_{0}^{\infty}\frac{\sqrt{\beta}dx}{\sqrt{x}(1+\beta
x)}\ln\left(\frac{N}{\alpha^{2}}\times\frac{(1+x)^{N-1}\prod\limits_{k=1}^{\left[\frac{N-1}{2}\right]}
\left|x-x_{k}^{B}\right|}{\prod\limits_{k=1}^{\left[\frac{N}{2}\right]}\left|x-x_{k}^{A}\right|^{3}}\right).
\]
 Now, making the following change of
variable $x=\frac{1}{\beta}\tan^{2}\frac{\Theta}{2}$, and taking
into account that degree of numerators and denominator are equal
for both even and odd values of N, we get:
\begin{gather*}
h(\mu,\Phi_{N}^{(\alpha)}(x))=\frac{1}{\pi}\int_{0}^{\infty}d\theta\Bigg\{\ln
\left(\frac{N}{\alpha^{2}}\right)+(N-1)\ln|\beta+1+(\beta-1)\cos\theta|\\
\qquad {}+\sum_{k=1}^{\left[\frac{N-1}{2}\right]}\!\ln
\left|1-x_{k}^{B}\beta+\left(1+x_{k}^{B}\beta\right)\cos\theta\right|
-3\sum_{k=1}^{\left[\frac{N}{2}\right]}\ln
\left|
1-x_{k}^{A}\beta+\left(1+x_{k}^{A}\beta\right)\cos\theta\right|\Bigg\}.
\end{gather*}
 Using the following integrals:
\[
\frac{1}{\pi}\int_{0}^{\pi}\ln|a+b\cos\theta|= \left\{
\begin{array}{ll}
\displaystyle \ln\left|\frac{a+\sqrt{a^{2}-b^{2}}}{2}\right|, \quad & |a| >| b|,
\vspace{2mm}\\
\displaystyle \ln\left|\frac{b}{2}\right|, & |a| \leq| b|,
\end{array}\right.
\]
we get:
\begin{equation}
 h\left(\mu,\Phi_{N}^{(\alpha)}(x)\right)=
\ln\left(\frac{N\left(1+\beta+2\sqrt{\beta}\right)^{N-1}}
{\left(\sum\limits_{k=0}^{\left[
\frac{N}{2}\right]}C_{2k}^{N}\beta^{k}\right)
\left(\sum\limits_{k=0}^{\left[
\frac{N-1}{2}\right]}C_{2k+1}^{N}\beta^{k}\right)}\right).
\end{equation}

\label{Jafarizadeh-lastpage}
\end{document}